\documentclass[reprint,aps,epsfig,superscriptaddress,twocolumn]{revtex4-1}
\usepackage{bm}
\usepackage{amsfonts}
\usepackage[dvips]{graphicx}
\usepackage{mathrsfs}
\usepackage[intlimits]{amsmath}
\usepackage{textcomp}
\usepackage[colorlinks, linkcolor=black, urlcolor=black, citecolor=black]{hyperref}
\renewcommand{\thefigure}{\arabic{figure}}
\newcommand{\ket}[1]{\vert #1 \rangle}

\begin{document}
	
\preprint{}

\newcommand{\thetitle}{Scalable and modular generation of multipartite entangled states through memory-enhanced fusion}

\title{\thetitle}

\author{Jixuan Shi}
\altaffiliation{These authors contributed equally to this work}
\affiliation{Center for Quantum Information, IIIS, Tsinghua University, Beijing 100084,
PR China}

\author{Sheng Zhang}
\altaffiliation{These authors contributed equally to this work}
\affiliation{Center for Quantum Information, IIIS, Tsinghua University, Beijing 100084,
PR China}

\author{Yukai Wu}
\altaffiliation{These authors contributed equally to this work}
\affiliation{Center for Quantum Information, IIIS, Tsinghua University, Beijing 100084,
PR China}
\affiliation{Hefei National Laboratory, Hefei 230088, PR China}

\author{Yuedong Sun}
\affiliation{Center for Quantum Information, IIIS, Tsinghua University, Beijing 100084,
PR China}

\author{Yibo Liang}
\affiliation{Center for Quantum Information, IIIS, Tsinghua University, Beijing 100084,
PR China}

\author{Hai Wang}
\email{wanghai@sxu.edu.cn}
\affiliation{The State Key Laboratory of Quantum Optics and Quantum Optics Devices,
Institute of Opto-Electronics, Shanxi University, Taiyuan 030006, China}
\affiliation{Collaborative Innovation Center of Extreme Optics, Shanxi University, Taiyuan 030006, China}

\author{Yunfei Pu}
\email{puyf@tsinghua.edu.cn}
\affiliation{Center for Quantum Information, IIIS, Tsinghua University, Beijing 100084,
PR China}
\affiliation{Hefei National Laboratory, Hefei 230088, PR China}

\author{Luming Duan}
\email{lmduan@tsinghua.edu.cn}
\affiliation{Center for Quantum Information, IIIS, Tsinghua University, Beijing 100084,
PR China}
\affiliation{Hefei National Laboratory, Hefei 230088, PR China}

\begin{abstract}
   Efficient generation of large-scale multipartite entangled states is a critical but challenging task in quantum information processing.
   Although generation of multipartite entanglement within a small set of individual qubits has been demonstrated, further scale-up in system size requires the connection of smaller entangled states into a larger state in a scalable and modular manner.
   Here we achieve this goal by implementing memory-enhanced fusion of two multipartite entangled states via photonic interconnects.
   Through asynchronous preparation of two tripartite W-state entanglements in two spatially-separated modules of atomic quantum memories and on-demand fusion via single-photon interference, we demonstrate the creation of a four-partite W-state entanglement shared by two remote quantum memory modules in a heralded way. We further transfer the W-state from the memory qubits to the  photonic qubits,
   and confirm the genuine four-partite entanglement through witness measurements.  We then demonstrate memory-enhanced scaling in efficiencies in the entanglement fusion.
   The demonstrated scalable generation and fusion of multipartite entangled states pave the way towards realization of large-scale distributed quantum information processing in the future.
\end{abstract}
\maketitle

\section{Introduction}

Multipartite entangled states, such as Greenberger-Horne-Zeilinger (GHZ) states~\cite{greenberger1990bell}, cluster states ~\cite{raussendorf2003measurement,nielsen2004optical}, and W states~\cite{dur2000three}, are fundamental resources for various quantum information processing tasks~\cite{knill2001scheme,markham2008graph,toth2012multipartite}. Among the multipartite entangled states, W states feature exceptional robustness against particle loss and have numerous quantum information applications, including secret sharing, dense coding, fault-tolerant teleportation, and distributed quantum metrology ~\cite{photonic_repeater,choi2010entanglement,shi2002teleportation,lipinska2018anonymous,agrawal2006perfect,Covey}. Significant progress has been achieved in generation of multipartite entangled state on various physical platforms~\cite{istrati2020sequential,yang2022sequential,thomas2022efficient,haffner2005scalable,solidstatewstate}. State-of-the-art demonstrations such as $51$-qubit superconducting cluster state~\cite{cao2023generation}, $32$-qubit trapped-ion GHZ state~\cite{moses2023race}, $14$-qubit photonic GHZ state~\cite{thomas2022efficient}, and $25$-qubit atomic-ensemble W state~\cite{pu2018experimental} have been achieved recently.

Despite these remarkable experimental achievements, the scale-up of a multipartite entangled state still remains a significant challenge. Quantum network connection of smaller modular architectures into bigger ones provides a promising method to overcome the scalability issue ~\cite{cirac1999distributed,cacciapuoti2019quantum,monroe2014large,thomas2024fusion,briegel1998quantum,Gisin,duan2001long}. The modular implementation includes intra-module entanglement generation and inter-module entanglement connection via photonic channels. For photonic systems, the local entanglement generation itself is inherently probabilistic, with the success probability decaying exponentially with the number of entangled qubits~\cite{yang2022sequential,thomas2022efficient}. In contrast, matter qubit systems such as trapped ions or superconducting circuits can prepare intra-module entanglement in a deterministic way. Nevertheless, their inter-module connection remains probabilistic due to the imperfect quantum interfaces, optical transmission losses, and photon detection inefficiencies. Therefore, both systems discussed above are subject to exponentially increasing costs with system size if a synchronous connection scheme without quantum memory is used. To resolve the scale-up problem, quantum-memory-enhanced schemes such as quantum repeater protocol are developed~\cite{briegel1998quantum,Gisin,duan2001long,pu2021experimental,kaneda2017quantum,langenfeld2021quantum,bhaskar2020experimental}. Due to the faithful quantum storage, the multipartite entangled states within each individual module can be generated in an asynchronous way and the photonic interconnections between modules are performed on demand. With memory enhancement, the efficiency of establishing multipartite entanglement can be significantly improved, which enables the efficient creation of a large-scale multipartite entangled state. So far, the asynchronous creation of more than one nonlocal multipartite entangled states distributed across individual photonic quantum interfaces has not been realized, let alone the on-demand fusion of multipartite entangled states via memory enhancement.

\begin{figure*}
\includegraphics[width=\linewidth]{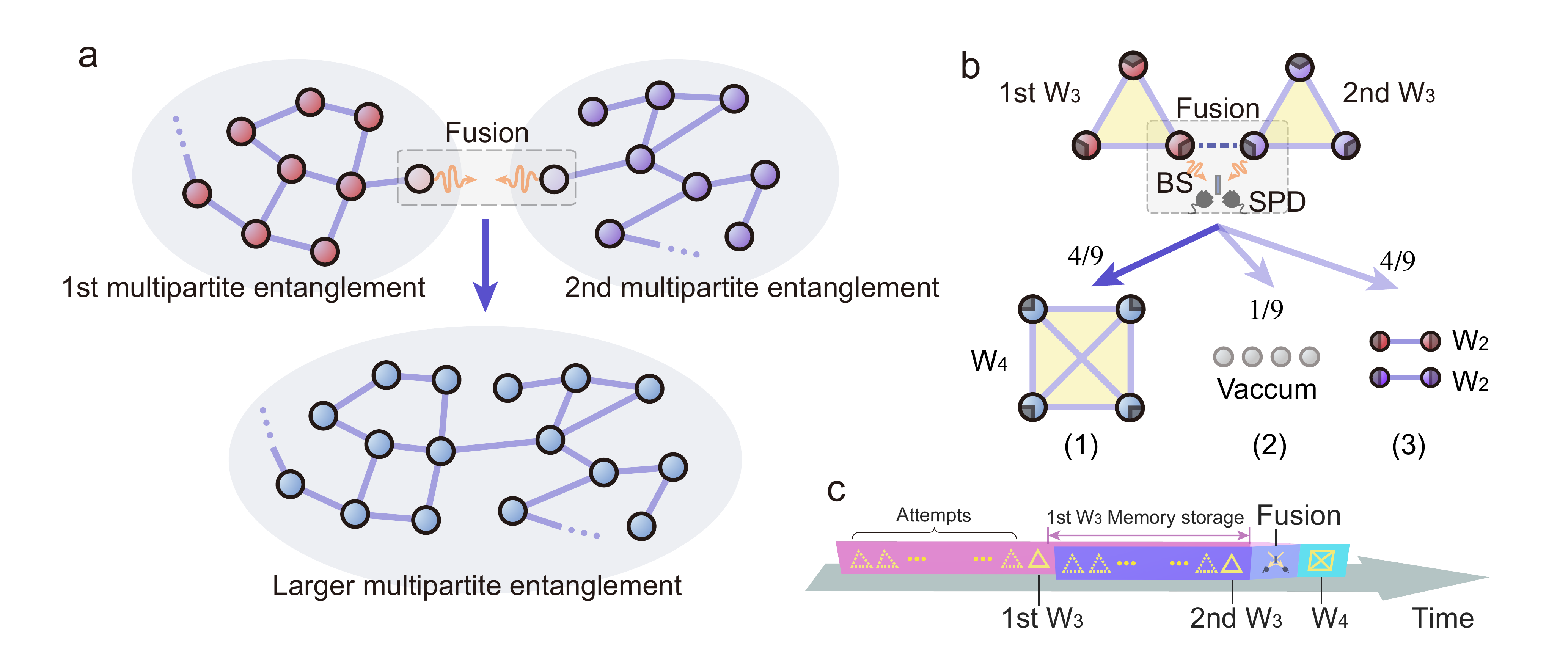}
\caption{
\textbf{Schematic of the experiment.}
\textbf{a}, Scale-up of a multipartite entangled state in a modular way under quantum network architecture. Local multipartite entangled states are generated within each individual module. Remote entanglement between modules is established by measurement-induced fusion with the help of quantum interfaces between matter and photonic qubits, which enables the creation of a larger non-local multipartite entanglement.
\textbf{b}, Fusion of two tripartite W-state entanglements. Six micro-ensembles are selected as matter qubit carriers from a two-dimensional atomic-ensemble quantum memory array and are divided into two modules. Tripartite W states are sequentially generated within each module. Then one micro-ensemble is chosen from each module, and the stored matter qubits are converted into photonic qubits for single-photon interference (fusion). There are three possible cases in the interference which correspond to three different states of the remaining four matter qubits: (1) successful generation of a four-partite W state; (2) four ensembles in the vacuum state without entanglement; (3) two separate bipartite entangled states.
\textbf{c}, The experimental sequence.
}\label{fig1}
\end{figure*}

\begin{figure}
  \centering
  \includegraphics[width=\linewidth]{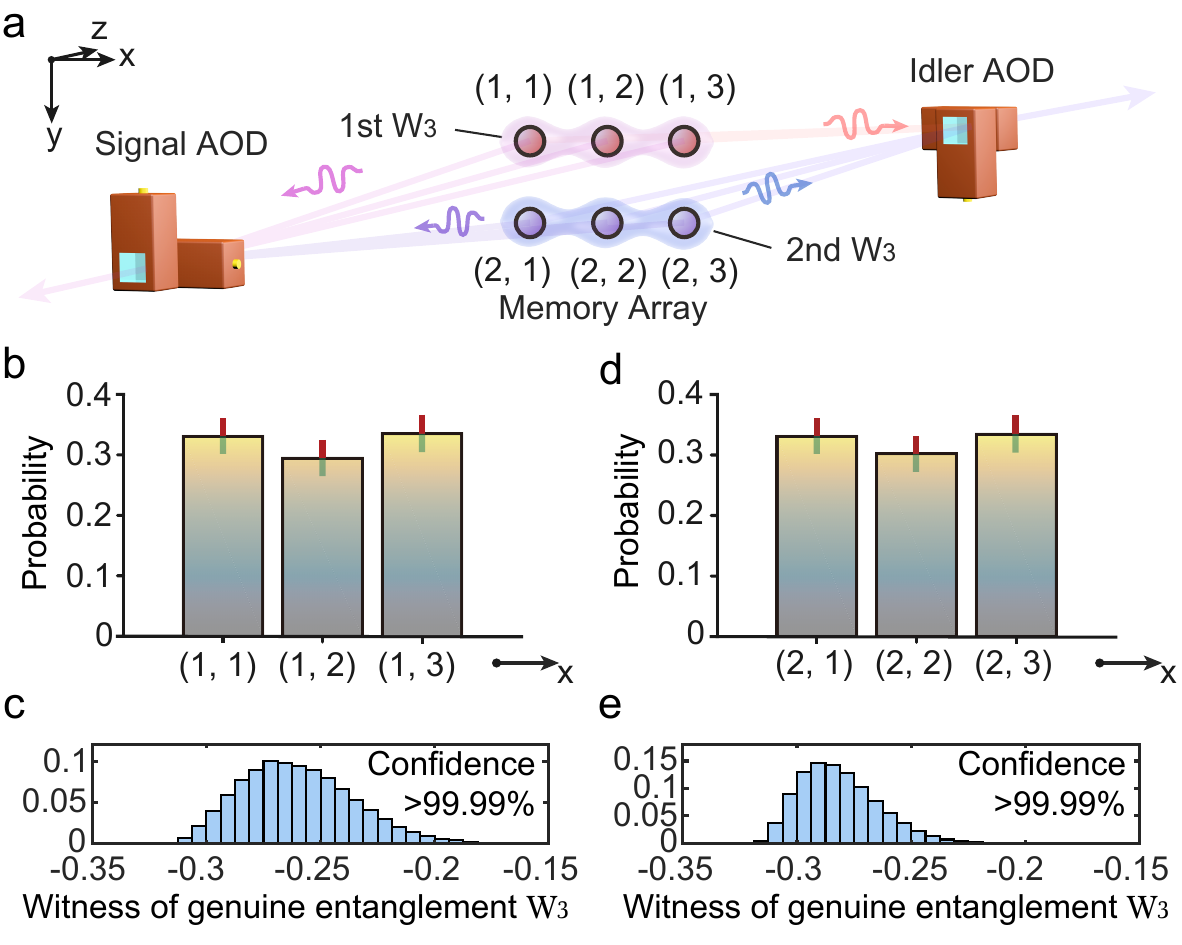}
  \caption{
\textbf{Characterization of two asynchronously prepared tripartite W-state entanglements.}
\textbf{a}, Schematic of generating and characterizing two $|W_3\rangle$ states in an array of atomic ensembles.
\textbf{b}, Population distribution of spin-wave excitations across the three micro-ensembles for the first  tripartite W state. The spatial coordinates of the ensembles are (1,1), (1,2), and (1,3). More detailed configuration is illustrated in
the Supplementary Information.
\textbf{c}, Entanglement witness distribution for the first tripartite W state. A negative value indicates genuine tripartite entanglement. Experimental data confirms that the witness value falls below zero with a probability over $99.99\%$.
\textbf{d,e}, Population and entanglement witness for the second tripartite W state.
  }\label{fig2}
\end{figure}

In this work, we demonstrate memory-enhanced fusion of two multipartite entangled states, as illustrated in Fig.~\ref{fig1}. We sequentially prepare two tripartite W-state entanglements in a heralded way, which are distributed in two spatially-separated regions in a two-dimensional array of long-lived quantum memories. Due to the long lifetime of the atomic-ensemble quantum memories, the first generated W-state entanglement is preserved coherently until the successful preparation of the second W-state entanglement. Upon the successful generation of both W-state entanglements, one of the memory qubits from each tripartite W state is converted into a photonic qubit, and these two photonic qubits are interferenced on a beamsplitter. A photon detection event heralds the successful fusion of these two tripartite W-state entanglements and yields a four-partite W-state entanglement distributed across four remote quantum memories shared by the two modules. By converting the spin-wave excitations in each quantum memory into single photons, we further create four-partite photonic W-state entanglement efficiently and verify the four-partite genuine entanglement. By decoupling the entanglement preparation and fusion processes through quantum memory, our protocol achieves a linear scaling in multipartite entanglement generation efficiency with respect to the tripartite W state generation probability, surpassing the quadratic scaling without memory enhancement. Our scheme in this experiment which scales up the size of multipartite entanglements via optical interconnects is scalable and can be further extended over multiple nodes in different experimental setups or even metropolitan-scale quantum network nodes. This work therefore provides a useful building block for generating large-scale nonlocal multipartite entangled states and can contribute to the realization of distributed quantum information processing in the future.

\section{Results}

Our experiment is based on a spatially multiplexed two-dimensional atomic quantum memory array, where individual addressing of micro-ensembles is achieved via acousto-optic deflectors (AODs)~\cite{pu2017experimental, lan}. Each micro-ensemble has a coherence time of about $500\,\mu$s through a collinear DLCZ (Duan-Lukin-Cirac-Zoller) scheme~\cite{zhang2024fast,zhang2024realization}. We first use a DLCZ protocol to generate two tripartite W-state entanglements distributed across two modules of micro-ensembles, respectively~\cite{pu2018experimental}. As shown in Fig.~\ref{fig2}, a write pulse is split into three spatially separated beams using a write AOD, enabling simultaneous excitation of a linear array of three micro-ensembles labeled as (1,1), (1,2), and (1,3). The signal photons emitted from these micro-ensembles are interferometrically combined on a signal AOD with the outputs coupled to a single-photon detector. A photon detection event heralds that the three micro-ensembles are in a tripartite W-state entanglement~\cite{pu2018experimental}. The resulted entangled state can be expressed as
\begin{equation}\label{W3}
  |W_3\rangle=\frac{1}{\sqrt{3}} (|100\rangle+|010\rangle+|001\rangle)
\end{equation}
where $|1\rangle\,(|0\rangle)$ denotes the presence (absence) of a collective spin-wave excitation in a micro-ensemble. The relative phase between each component is actively compensated by tuning the phase of the RF signals driving the AOD.

\begin{figure}
  \centering
  \includegraphics[width=\linewidth]{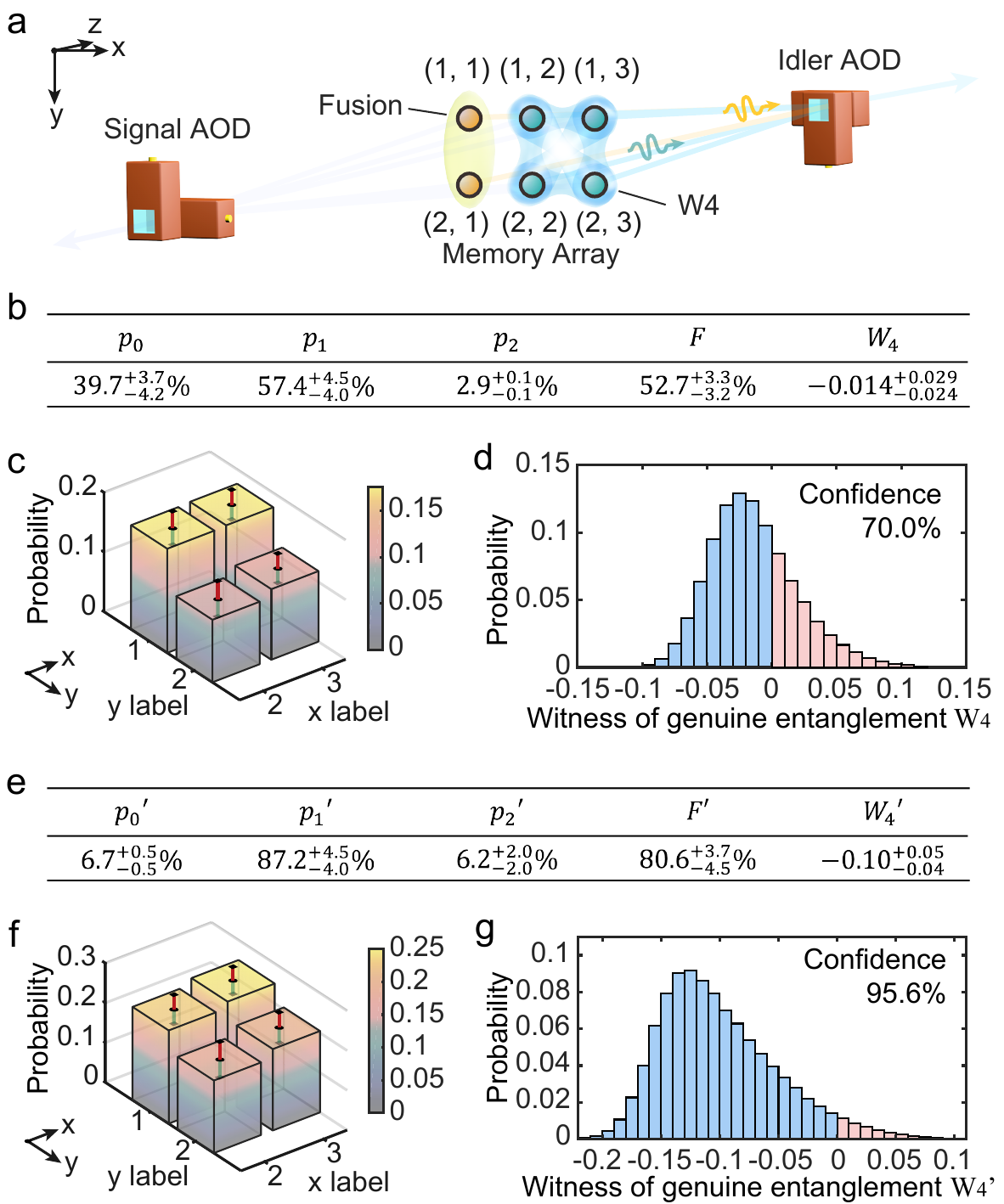}
  \caption{
  \textbf{Entanglement fusion and the verification of genuine four-partite entanglement.}
  \textbf{a}, Schematic of the entanglement fusion and genuine entanglement verification.
  \textbf{b}, The measured value for population in zero-, one- and two-excitation subspaces  $p_0,\,p_1,\,p_2,$ the fidelity $F$, and the entanglement witness $W_4$ of the four remaining micro-ensembles. Optimal witness in the form of $\mathcal{W}_k=\alpha_k P_0+\beta_k P_1+\gamma_k P_2-|W_N\rangle \langle W_N |$ are given by $\alpha=0.0977$, $\beta=0.7775$, $\gamma=1$.
\textbf{c}, Population distribution of spin-wave excitations across the four micro-ensembles for the four-partite W state.
\textbf{d}, The distribution of entanglement witness $W_4$. The probability of genuine four-partite entanglement verified by $\text{tr}[\mathcal{W}_4 \rho_e]<0$ is $70.0\%$.
\textbf{e-g}, The measured fidelity, population, and entanglement witness for the photonic four-partite W state after post-selection of successful photon detection.  Optimal witness parameters are $\alpha^\prime=0.3854$, $\beta^\prime=0.7525$, $\gamma^\prime=0.4101$. The probability of achieving a genuine four-partite entanglement is $95.6\%$ from these measurements.
  }\label{fig3}
\end{figure}

Due to the probabilistic nature of the DLCZ protocol, the generation of the tripartite W states needs to be repeated for many times. Upon recording the first heralding signal, we reconfigure the write AOD to address the next row of three micro-ensembles labeled as (2,1), (2,2), (2,3) and repeat the excitation sequence until a second photon detection event heralds the generation of another tripartite W-state entanglement. Subsequently, the collective excitation modes stored in micro-ensembles (1,1) and (2,1) are coherently converted into idler photonic modes. These photonic modes are combined into the idler AOD which acts as a beamsplitter (BS)~\cite{pu2018experimental} for the entanglement fusion operation, as shown in Fig.~\ref{fig3}. A photon detection event on the idler detector projects the four remaining micro-ensembles into a four-partite W state. Finally, the four-partite entangled state stored in the atomic ensemble array is read out into a photonic W state, with its entanglement properties verified through an entanglement witness.

As depicted in Fig.~\ref{fig1}b, there are three possible cases in the entanglement fusion depending on the number of spin-wave excitations in the selected two ensembles for the fusion operation. These cases are: (1) Exactly one ensemble has a spin-wave excitation. This accounts for $4/9$ of all the possible cases, and a single-photon detection event heralds the successful creation of a four-partite W state across the remaining ensembles; (2) Both ensembles contain an excitation inside. This happens with a probability of $1/9$, and a photon click leaves the remaining system in a vacuum state; (3) Neither ensemble contains an excitation. This has a probability of $4/9$, and the fusion process yields no photon click. In this case the remaining ensembles are in separable bipartite entangled states. Assuming the overall photon retrieval and detection efficiency in our experiment is $\eta$, the two situations with photon detections (case (1) and (2)) have success probabilities of $4/9\times\eta\times0.5=2\eta/9$ (here $0.5$ is due to the $50$:$50$ beamsplitter) and $1/9\times\eta\times2\times0.5=\eta/9$ (here $2$ is due to the doubled photon detection probability with double excitations), respectively. The total success probability of receiving a photon detection click in the fusion is thus $2\eta/9+\eta/9=\eta/3$. As we cannot distinguish these two cases, a successful fusion results in a mixed state of $\rho=\frac{2}{3}|W_4\rangle\langle W_4|+\frac{1}{3}\rho_{\text{vac}}$ ($\rho_{\text{vac}}$ denotes vacuum state). Nevertheless, this vacuum component does not influence the utilization of the created four-partite W state as this part can be post-selected by photon detection events during the entanglement verification step, fusion with other W-states, or further applications via photon measurement.

\begin{figure}
  \centering
  \includegraphics[width=0.9\linewidth]{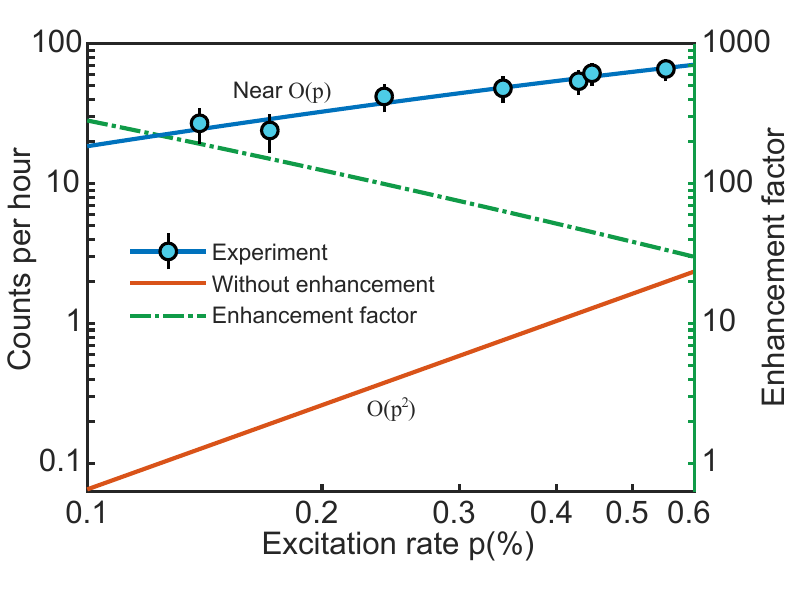}
  \caption{
  \textbf{Memory-enhanced scaling in generation efficiency of the four-partite W-states.}
Four-partite entanglement generation efficiency as a function of the success probability $p$ for preparing a tripartite entangled state in each attempt. The blue dots are the measured four-photon coincidence counts per hour, including two signal photons heralding the two successfully prepared tripartite W-states, one heralding photon for the fusion operation, and one readout photon for entanglement verification. The blue solid line denotes the simulated four-photon generation rate in this experiment, while the orange solid line represents the theoretical prediction for a protocol without memory enhancement. The green dotted line shows the simulated enhancement factor between the memory-enhanced and memory-less cases. Error bars represent one standard deviation.
  }\label{fig4}
\end{figure}


We first verify multipartite quantum entanglement in the two tripartite W states. The experimentally generated states $\rho_e$ can be expressed as $\rho_e=p_0\rho_0+p_1\rho_1+p_2\rho_2$, where $p_0,p_1,p_2$ and $\rho_0,\rho_1,\rho_2$ denote the population and  corresponding density matrix with zero, one, and double excitations in the spin-wave modes, respectively. The fidelity of the prepared state is defined as $F=\langle W_3|\rho_e |W_3\rangle$. To verify tripartite entanglement, we use an entanglement witness to lower bound the entanglement depth $k\,(k\le N)$, which means the state has at least genuine k-partite entanglement. Here we use an entanglement witness written as $\mathcal{W}_k=\alpha_k P_0+\beta_k P_1+\gamma_k P_2-|W_N\rangle \langle W_N |$, where $P_n\,(n=0,1,2)$ represents projectors onto the subspace with $n$ excitations in the spin-wave modes and the parameters $\alpha_k,\,\beta_k,\,\gamma_k\geq 0$ are numerically optimized to guarantee $\text{tr}[\mathcal{W}_k\rho] \geq0$  for any state $\rho$ with entanglement depth less than $k$. Thus $\text{tr}[\mathcal{W}_3 \rho_e]<0$ serves as a sufficient condition to verify the presence of at least tripartite genuine entanglement among the three micro-ensembles~\cite{pu2018experimental,guhne2009entanglement}.

To bound the entanglement depth, we experimentally measure the fidelity $F$ and the population $p_0,\,p_1,\,p_2$. The detailed measurement procedure is explained in the Supplementary Information. We use an excitation probability of $p=0.15\%$ to generate the tripartite W-state, and obtain the population distribution  as shown in Fig.~\ref{fig2}b and Fig.~\ref{fig2}d. In Fig.~\ref{fig2}c and \ref{fig2}e, we present the distribution of the entanglement witness $\text{tr}[\mathcal{W}_3\rho_e]$.  A genuine tripartite entanglement can be verified with a confidence level over $99.99\%$ in both of the two tripartite entangled states generated asynchronously.

Following the successful entanglement fusion heralded by a detector click, we characterize the generated four-partite W state via entanglement witness in a similar way. The measured population distribution $p_0,\,p_1,\,p_2$ and fidelity $F$ are presented in Fig.~\ref{fig3}b. The optimized entanglement witness yields $\text{tr}[\mathcal{W}_4 \rho_e]<0$ with a probability of $70.0\%$, which confirms that we generate a genuine four-partite entanglement among four distinct quantum memories with a high probability. The measured vacuum population $p_0=39.7\%$ is in good agreement with the theoretical value $1/3$. In addition, we further convert the spin-wave entangled state into a photonic multipartite entanglement, and the vacuum part can be largely post-selected by the photon detection events ($p_0'=6.7\%$, see Fig.~\ref{fig3}e. $p_0'$ is not zero due to the background noise). With the vacuum part mostly removed, the genuine four-partite photonic entanglement can be yielded with a probability of $95.6\%$, as shown in Fig.~\ref{fig3}g. This reveals that the inherent vacuum part appeared in the fusion operation does not influence the quality of the fused W state, similar to the case in the DLCZ quantum repeater protocol~\cite{duan2001long}.

Finally, we emphasize that our protocol realizes a memory-enhanced efficiency in the fusion of two multipartite entangled states. By measuring the four-photon coincidence rate (including two heralding signal photons during the two tripartite W-states generation, one heralding idler photon from the fusion operation, and one idler photon for verifying the final W state), we observe a linear relationship $\mathcal{O}(p)$ between the success rate of the four-partite entanglement by fusion and the success probability of each tripartite entanglement $p$, as shown in Fig.~\ref{fig4}. The experimental data is in good agreement with the theoretical predictions (blue dots and curve in Fig.~\ref{fig4}). In contrast, a memory-less protocol would require simultaneous generation of two tripartite entangled states with a joint probability of $p^2$, leading to a quadratic scaling $\mathcal{O}(p^2)$ (orange curve in Fig.~\ref{fig4}). The transition from quadratic scaling to linear scaling in efficiency clearly shows this quantum-memory-enhanced scheme is much more advantageous over the memory-less scheme. This memory-enhanced scheme, which features a quantum-repeater-like efficiency, paves the way to efficiently generate large-scale multipartite entanglement in the future.

\section{Discussion}

In summary, We demonstrate the entanglement fusion of two multipartite entangled states distributed in different quantum memory modules into a larger multipartite entangled state through a photonic link. The use of quantum memory improves the fusion efficiency from a quadratic scaling to a linear scaling. Besides, due to atomic quantum interfaces used in this experiment, we can convert the memory-based W state to photonic W state in an efficient way. Our experiment therefore demonstrates a way to efficiently generate larger scale multipartite entangled states in a modular fashion. This modular approach of scaling up the size of a quantum information system via photonic interfaces can also be applied to other physical platforms such as trapped ion~\cite{saha2025high,main2025distributed} and superconducting systems~\cite{mirhosseini2020superconducting,almanakly2025deterministic}. In the future, we can further convert the photons to the telecom band to interconnect multipartite entanglements distributed in quantum network nodes separated by metropolitan distances~\cite{zhang2024fast,Ikuta,chang}. This would enable the creation of nonlocal multipartite quantum entanglement states distributed over a large area, holding significant practical value for both fundamental investigations and various quantum information applications, such as distributed quantum computing and sensing, quantum secret sharing, and the future quantum internet~\cite{mermin1990extreme,Kimble,Internet}.


\noindent\textbf{Data availability} The data that support the findings of this study are available from the
corresponding authors upon request.

\noindent\textbf{Acknowledgements} This work is supported by Innovation Program for Quantum Science and Technology (No.2021ZD0301102), the Tsinghua University Initiative Scientific Research Program, the Ministry of Education of China through its fund to the IIIS, and National Key Research and Development Program of China (2020YFA0309500). Y.P. acknowledges support from the Dushi Program from Tsinghua University.

\noindent\textbf{Author Information} The authors declare no competing financial interests. Correspondence and requests for materials should be addressed to H.W. (wanghai@sxu.edu.cn), Y.P. (puyf@tsinghua.edu.cn), or L.D. (lmduan@tsinghua.edu.cn).

\noindent\textbf{Competing Interests} The authors declare no competing financial interests.

\onecolumngrid


\setcounter{equation}{0}
\setcounter{figure}{0}
\setcounter{table}{0}
\setcounter{page}{1}
\setcounter{section}{0}
\makeatletter
\renewcommand{\theequation}{S\arabic{equation}}
\renewcommand{\thefigure}{S\arabic{figure}}
\renewcommand{\thetable}{S\arabic{table}}

\pagebreak
\begin{center}
\Large Supplemental Information for\\\textbf{``Scalable and modular generation of multipartite entangled states through memory-enhanced fusion"}
\end{center}
	
\section{Experimental setup and quantum memory performances}

In this section, we illustrate the experimental setup and the corresponding memory performances in detail. Fig.~S1a demonstrates our experimental setup, which consists of a $2\times 3$ array of micro-ensembles of cold $^{87}$Rb atoms. Each micro-ensemble functions as an individual quantum interface and is addressed by four pairs of AODs, as shown in Fig.~S1b. To generate matter-photon entanglement at each quantum interface, we employ the DLCZ protocol (energy levels are shown in Fig.~S1c). Heralded by a signal photon detection event, three quantum interfaces are projected onto a multipartite entangled state $|W_3\rangle$. In order to prolong the coherence time of each quantum interface, both the write and read beams are configured collinearly with the optical path used for collecting signal and idler photons via $90$:$10$ beamsplitters. Additionally, three etalons are incorporated into each photon path to effectively suppress the leakage of the write or read beam with a high extinction ratio over $80\,$dB. Fig. S1c and S1d present the measured end-to-end memory efficiency and the cross-correlation function across all six interfaces, respectively. The average coherence time across all memory cells is 428$\,\mu$s.

\begin{figure*}
  \centering
  \includegraphics[width=\linewidth]{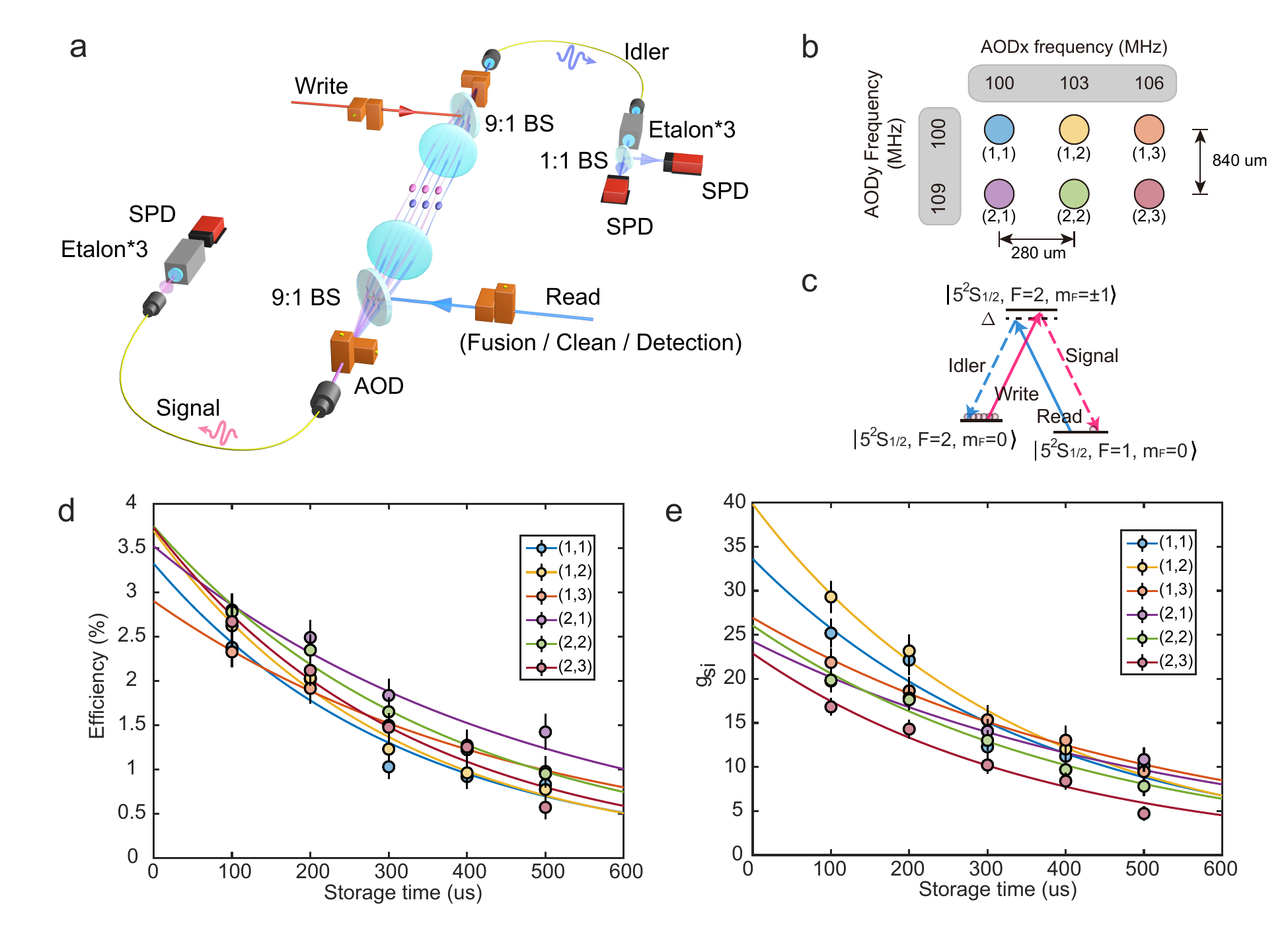}
  \caption{\textbf{Experimental configuration and quantum memory performances.}  \textbf{a}, The experimental setup. \textbf{b}, AOD frequencies for addressing each micro-ensemble and the spatial separations. \textbf{c}, The energy levels used in the experiment. We use the DLCZ protocol to generate atom-photon entanglement. \textbf{d}, Measured end-to-end retrieval efficiency of all the $6$ memory cells with optical transmission losses and detector inefficiencies included.  \textbf{e}, Measured cross-correlation function of all the $6$ memory cells.}
\end{figure*}

\section{Control sequence}

\begin{figure}
  \centering
  \includegraphics[width=\linewidth]{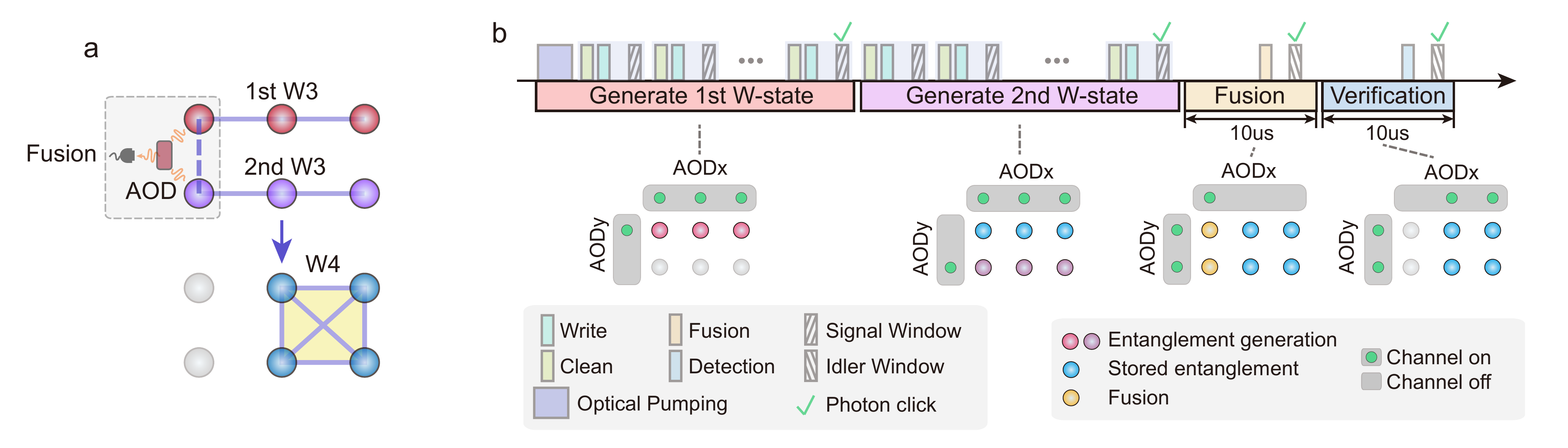}
  \caption{\textbf{Experimental sequence}. \textbf{a}, Scheme of W state fusion on the $2\times3$ interface array. \textbf{b}, Experimental sequence.}
\end{figure}

In this section, we demonstrate the protocol and sequence used in our experiment. Fig.~S2a shows an overview of the  experiment. The first tripartite W state is generated in the upper three interfaces, while the second W state is prepared in three interfaces on the lower row. To connect these two W states, the leftmost cells in these two rows are read out and interfered on an AOD-based beamsplitter. Upon the detection event of an idler photon, the remaining four cells (1,2), (1,3), (2,2), and (2,3) are projected onto a four-particle W state in a heralded way. The detailed control sequence is depicted in Fig.~S2b. Initially, atomic ensembles are prepared in the ground state $\ket{F=2, m_F=0}$ via optical pumping and the three cells on the upper row are addressed by activating the first Y channel of the AODs. All three X channels are simultaneously opened to excite these three cells at the same time and measure the signal photon modes in a superposition basis. These first three cells are excited repeatedly until a signal photon is detected. If no signal photon is detected, a cleaning pulse is applied and the excitations starts over. Each attempt costs $650\,$ns in total. Once a signal photon on the superposition basis is detected, the first three interfaces are projected onto a tripartite W state. We then shift the Y channel of all the four AODs to the second row, and the second W state is generated using the same method. Here we limit the maximum number of excitation attempts on the second row to $300$ to prevent the readout efficiency of the first W state from being too low. After the preparation of both tripartite W states, we read out the two cells in the leftmost column by a read pulse. The idler photons from the two cells are interfered on the idler AOD which serves as a $50$:$50$ beamsplitter. If a successful fusion is heralded by an idler photon detection event, we bound the genuine entanglement of the remaining four cells by (1) addressing each cell individually to detect the excitation population and (2) reading all the interfaces out at the same time and measure the entanglement fidelity on the W state basis.

\begin{figure}
  \centering
  \includegraphics[width=\linewidth]{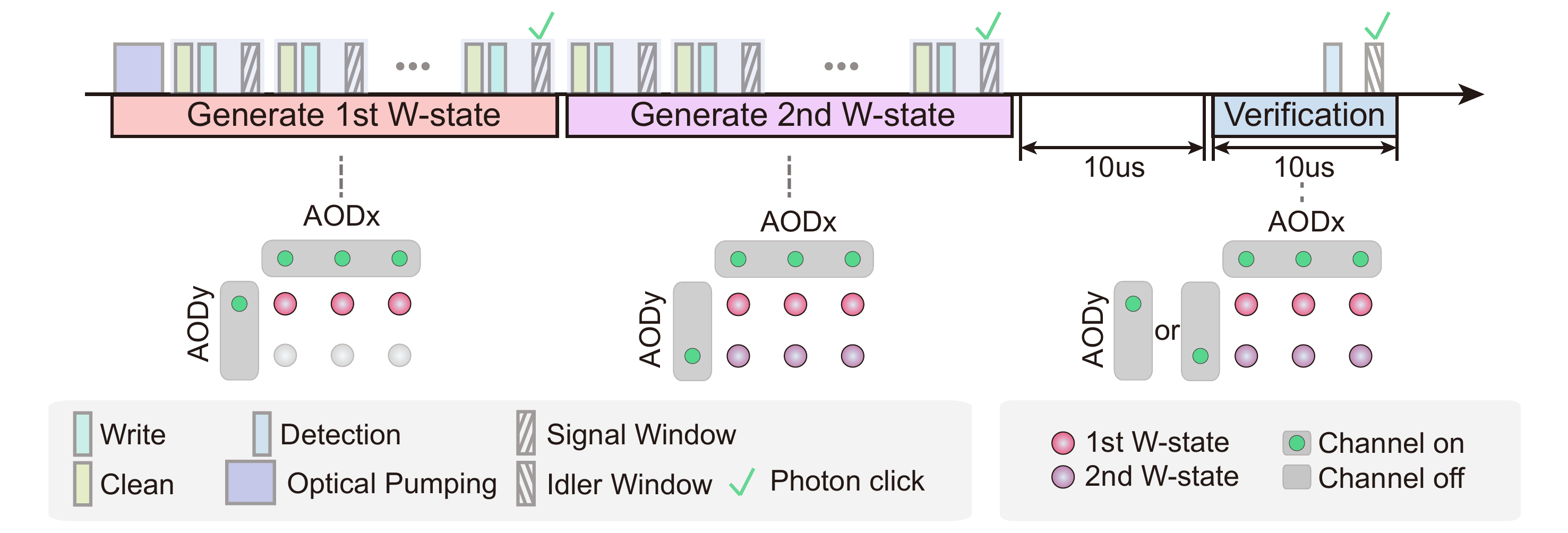}
  \caption{\textbf{Control sequence for the verification of tripartite W states}. The experimental sequence for generating and verifying two tripartite W states individually.}
\end{figure}

When we need to verify the two tripartite W states before fusion, the  corresponding sequence is shown as Fig.~S3. Here the fusion operaton is not applied and we open different AOD channels to verify different tripartite W states.

\section{Entanglement witness}
To verify the genuine multipartite entanglement in the experimentally prepared W-state, we use an entanglement witness in the form of \cite{pu2018experimental,guhne2009entanglement}
\begin{equation}
\mathcal{W}_{k}=\alpha_{k}P_{0}+\beta_{k}P_{1}+\gamma_{k}P_{2} - |W_{N}\rangle \langle W_{N}| \label{eq:witness},
\end{equation}%
where $P_{n}$ ($n=0,1,2$) represents projectors on the $n$-excitation subspace and $|W_{N}\rangle$ is the $N$-qubit W state.
To ensure that $\mathcal{W}_{k}$ makes a valid entanglement witness for genuine $k$-partite entanglement, we choose the non-negative parameters $\alpha_k$, $\beta_k$ and $\gamma_k$ such that $\langle\mathcal{W}_{k}\rangle = \mathrm{Tr}[\rho \mathcal{W}_{k}] \ge 0$ for any state $\rho$ that can be decomposed into the mixture of product states with less-than-$k$-partite entanglement. On the other hand, for the given W state $\rho_e$ prepared in the experiment with the experimentally measured populations $p_0=\mathrm{Tr}[\rho_e P_0]$, $p_1=\mathrm{Tr}[\rho_e P_1]$, $p_2=\mathrm{Tr}[\rho_e P_2]$, and fidelity $F=\langle W_{N}|\rho_e |W_{N}\rangle$, we optimize the valid witness parameters $\alpha_k$, $\beta_k$ and $\gamma_k$ to make $\mathrm{Tr}[\rho_e \mathcal{W}_k]=\alpha_k p_0 + \beta_k p_1 + \gamma_k p_2 - F$ as small as possible, so as to best certify the multipartite entanglement.

Below we summarize our algorithm to optimize these witness parameters in Ref.~\cite{pu2018experimental} for completeness. Since we are interested in W states close to the ideal $|W_{N}\rangle$ with genuine $N$-partite entanglement, we assume $k>2N/3$ which holds for the $(N=3, k=3)$ and $(N=4, k=4)$ cases considered in this work. In this scenario, for pure states without genuine $k$-partite entanglement, we can divide them into the tensor product of two components with the sizes of $l$ and $N-l$ ($l<k$ and $N-l<k$), respectively. We do not need to worry about dividing into three or more parts because in such situations we can still combine the smaller parts together and finally obtain two parts with the larger one smaller than $k$. Furthermore, owing to the permutation symmetry of $\mathcal{W}_{k}$ in Eq.~(\ref{eq:witness}), we only need to consider tensor product states in the form of $|a\rangle _{1,\cdots ,l}|b\rangle
_{l+1,\cdots ,N}$ where $N-l\le l < k$.

The pure state $|a\rangle _{1,\cdots ,l}$ (and similarly $|b\rangle
_{l+1,\cdots ,N}$) can generally be decomposed into $|a\rangle _{1,\cdots ,l}=a_{0}|g\rangle _{1,\cdots ,l}+a_{1}|e_{1}\rangle
_{1,\cdots ,l}+\cdots +a_{l}|e_{l}\rangle _{1,\cdots ,l}$ where $|g\rangle_{1,\cdots ,l}=|00\cdots 0\rangle_{1,\cdots ,l}$ and $|e_{r}\rangle_{1,\cdots ,l}\propto P_{r}|a\rangle _{1,\cdots ,l}$ is a
normalized state with exactly $r$ excitations. To ensure that a given set of non-negative witness parameters $\alpha_k$, $\beta_k$ and $\gamma_k$ always give non-negative $\langle\mathcal{W}_{k}\rangle$ for this type of tensor product states, we only need to check states in the form of $|a\rangle _{1,\cdots ,l}=a_{0}|g\rangle _{1,\cdots ,l}+a_{1}|W_l\rangle_{1,\cdots ,l}$ and $|b\rangle_{l+1,\cdots ,N}=b_{0}|g\rangle _{l+1,\cdots ,N}+b_{1}|W_{N-l}\rangle_{l+1,\cdots ,N}$ because the other discarded terms just further increase the positive values of $p_1$ and $p_2$ in the witness.

Now we can rewrite $a_{0}=\cos \theta _{1}$, $a_{1}=\sin \theta _{1}$, $b_{0}=\cos \theta_{2}$, $b_{1}=\sin \theta _{2}$ ($0\leq \theta _{1},\theta _{2}\leq \pi /2$), and express $\langle\mathcal{W}_{k}\rangle$ as
\begin{align}
f(\theta_1,\theta_2)=& \frac{1}{4}\bigg[\alpha _{k}(1+\cos 2\theta _{1})(1+\cos 2\theta
_{2})+2\beta _{k}(1-\cos 2\theta _{1}\cos 2\theta _{2})  \notag \\
& +\gamma _{k}(1-\cos 2\theta _{1})(1-\cos 2\theta _{2})-(1+\cos 2\theta
_{1})(1-\cos 2\theta _{2})  \notag \\
& +\frac{2l}{N}(\cos 2\theta _{1}-\cos 2\theta _{2})-\frac{2\sqrt{l(N-l)}}{N}%
\sin 2\theta _{1}\sin 2\theta _{2}\bigg].
\end{align}
Our target is to find valid parameters $\alpha_k$, $\beta_k$ and $\gamma_k$ such that $\langle\mathcal{W}_{k}\rangle$ takes non-negative values for any $0\leq \theta _{1},\theta _{2}\leq \pi /2$. This can be checked by numerically minimizing $f(\theta_1,\theta_2)$ for given $\alpha_k$, $\beta_k$ and $\gamma_k$. We can now scan $\alpha_k$, $\beta_k$ and $\gamma_k$ within $[0,1]$ and make a numerical array of valid witness parameters. Finally, for the given experimental data $p_0$, $p_1$, $p_2$ and $F$, we choose the optimal set of witness parameters to give the most negative expectation value $\langle\mathcal{W}_{k}\rangle$ as the optimal entanglement witness.

\section{General results of entanglement fusion via single photon interference}

\begin{figure}
  \centering
  \includegraphics[width= 0.8 \linewidth]{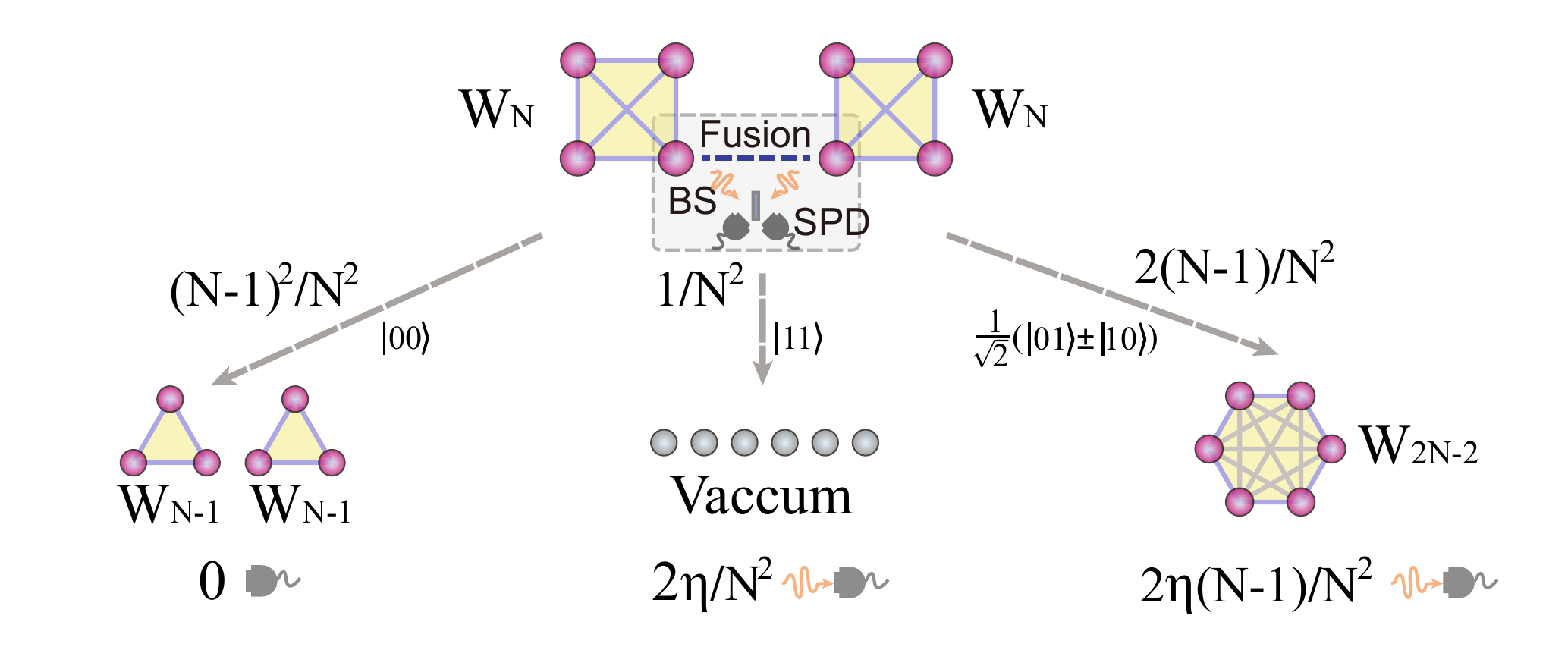}
  \caption{\textbf{Different cases for entanglement fusion}. Here we demonstrate the three possible cases in the entanglement fusion operation.}
\end{figure}

In this section, we derive the general outcomes of the fusion operation based on single-photon interference in our experiment. Assume we have prepared  two N-partite W states, which can be expressed as
\begin{equation}
\ket{W_N}=\frac{1}{\sqrt{N}}\sum_{j=1}^{N}\ket{0 \cdot\cdot\cdot 1_j \cdot\cdot\cdot 0}
\end{equation}
Here all the phases are set to zero for simplicity. During the fusion process, one memory mode from each W state is read out and the two photonic modes are interfered on a $50$:$50$ beam splitter. To derive the results of this fusion process, we can express the $N$-partite W state as
\begin{equation}
\ket{W_N}=\frac{1}{\sqrt{N}}\ket{1}\ket{Vac_{N-1}}+\frac{\sqrt{N-1}}{\sqrt{N}}\ket{0}\ket{W_{N-1}}
\end{equation}
where $|1\rangle$ or $|0\rangle$ represents the existence or not of an excitation in the first mode $j=1$ in the $N$-partite W state (the mode to be read out in the fusion), and $|W_{N-1}\rangle$ or $|Vac_{N-1}\rangle$ denotes the existence of an $(N-1)$-partite W state or a vaccum state in the remaining $N-1$ modes ($j=2\,\text{to}\,N$).  The initial product state can then be expanded as
\begin{align}
\ket{W_N}_A\ket{W_N}_B=&\frac{N-1}{N}\ket{0}_A\ket{0}_B\ket{W_{N-1}}_A\ket{W_{N-1}}_B+\frac{1}{N}\ket{1}_A\ket{1}_B\ket{Vac_{N-1}}_A\ket{Vac_{N-1}}_B \notag\\
&+\frac{\sqrt{N-1}}{N}\frac{1}{\sqrt{2}}(\ket{0}_A\ket{1}_B+\ket{1}_A\ket{0}_B)\ket{W_{2N-2}}^{+}_{AB}+\frac{\sqrt{N-1}}{N}\frac{1}{\sqrt{2}}(\ket{0}_A\ket{1}_B-\ket{1}_A\ket{0}_B)\ket{W_{2N-2}}^{-}_{AB}
\end{align}
where
\begin{align}
\ket{W_{2N-2}}^{+}_{AB} =&\frac{1}{\sqrt{2}}(\ket{W_{N-1}}_A\ket{Vac_{N-1}}_B+\ket{Vac_{N-1}}_A\ket{W_{N-1}}_B) \\
\ket{W_{2N-2}}^{-}_{AB} =&\frac{1}{\sqrt{2}}(\ket{W_{N-1}}_A\ket{Vac_{N-1}}_B-\ket{Vac_{N-1}}_A\ket{W_{N-1}}_B)
\end{align}

We can find that the fusion operation yields three possible outcomes as shown in Fig.~S4. The first case is that neither of the memory modes for the fusion contains an excitation, which can be expressed as $\ket{0}_A\ket{0}_B$ in Eq.~S5. This leads to two separated $(N-1)$-partite W states $\ket{W_{N-1}}_A\ket{W_{N-1}}_B$.  This case happens with a probability of $p=(\frac{N-1}{N})^2$. The second case is that both memory modes contain an excitation, which corresponds to the $\ket{1}_A\ket{1}_B$ item in Eq.~S5, leaves all the remaining qubits in a vacuum state $\ket{Vac_{N-1}}_A\ket{Vac_{N-1}}_B$. This case occurs with a probability of $p=\frac{1}{N^2}$. The third case is that there exist only one excitation in either two qubits for the fusion, which corresponds to $\frac{1}{\sqrt{2}}(\ket{01}\pm\ket{10})$ in Eq.~S5. The remaining $2N-2$ qubits are projected into a ($2N-2$)-partite W state of either $\ket{W_{2N-2}}^{+}_{AB}$ or $\ket{W_{2N-2}}^{-}_{AB}$, conditioned on which port the photon leaves the $50$:$50$ beamsplitter after the interference. This case happens with a probability of $p=\frac{2(N-1)}{N^2}$.

As discussed above, only the second and third cases will yield photon detection events but the first case will not. Thus conditioned on a photon detection event heralding the successful fusion, the remaining $2N-2$ qubits are in a mixed state of the ($2N-2$)-partite W state and an unwanted vacuum state. Assuming the total detection efficiency of the read out is $\eta$, the success probability of the second case is $p' = \frac{2\eta}{N^2}$ (in our case that $\eta$ is significantly lower than unity, the photon detection probability is roughly doubled for the double excitation) and the success probability of the third case is $p' = \frac{2(N-1)\eta}{N^2}$. Therefore, the total success probability of the fusion is $p' = \frac{2\eta}{N}$, and yields a $\frac{1}{N}$ vacuum part in the state after fusion. In our current experiment, we only collect and detect one of the two outputs of the $50$:$50$ beamsplitter, resulting in a total probability of $p' = \frac{\eta}{N}$ instead of $p' = \frac{2\eta}{N}$.

\end{document}